# FABRICATION AND CHARACTERIZATION OF A MICRO-BOILING METER FOR HEAT TRANSFER MEASUREMENTS ON A SINGLE NUCLEATION SITE

**H. Kharkwal[1*], S.K.H. Shah[2], L. Robert[1], J.M. Cote[3], E. Gavignet[2], S. Euphrasie[1], F. Lanzetta[2], L. Tadrist[4], M. Barthès[1]**

[1]Université Marie et Louis Pasteur, CNRS, institut FEMTO-ST, F-25000 Besançon, France
[2]Université Marie et Louis Pasteur, CNRS, institut FEMTO-ST, F-90000 Belfort, France
[3]Université Marie et Louis Pasteur, SUPMICROTECH, CNRS, institut FEMTO-ST, F-25000 Besançon, France
[4]Université d'Aix-Marseille, CNRS, IUSTI, Marseille, France

## ABSTRACT

Accurate characterization of heat transfer during pool boiling requires sensors capable of resolving rapid transients and highly localized thermal events. Conventional approaches often suffer from limited spatial resolution and slow thermal response due to intermediate material layers. To overcome these limitations and obtain accurate measurements of surface temperature and heat flux during pool boiling studies, we used a boiling meter. Two of these consist of commercially available sensors. To improve the study of vapor bubble dynamics, a third boiling meter was designed, dimensioned, and microfabricated. It was fabricated on a Borofloat 33–silicon composite substrate featuring a suspended Borofloat membrane with a central 200 µm micropillar and a laser-ablated nucleation site (100 µm × 115 µm). Two platinum resistance temperature detectors (Pt-RTDs) were integrated on both sides of the substrate to enable direct, localized, and time-resolved measurements of wall temperature and heat flux during bubble formation. Characterization of the microfabricated boiling meter was carried out in controlled radiative environment and in pool boiling. The microfabricated device exhibited a thermal characteristic time of around 200 ms in radiative tests (more than 30 times shorter under similar operating conditions than the times found with boiling meters composed of commercial sensors), but was able to capture a bubble frequency of 26 Hz, which validates its ability to resolve transient thermal events. Its suspended glass-membrane structure, combined with the direct liquid–RTD interface, provides both high temporal resolution and spatial sensitivity, essential for capturing the short timescales of bubble nucleation, growth, and departure.

## 1. INTRODUCTION

Boiling is an efficient form of heat transfer. Therefore, understanding boiling heat transfer is very important for designing high-performance heat transfer systems for use in areas such as power generation and electronics cooling. Despite several years of study, the process is still not fully understood, largely due to its complex behaviour and dependence on working conditions. Consequently, the investigation of the initial boiling events (onset of nucleate boiling, ONB) at a solitary nucleation site can yield invaluable insights into the underlying physics that governs boiling heat transfer. To achieve this, it is essential to accurately measure the local temperature and heat flux in order to study such a transient, spatially localised phenomenon.

There are numerous techniques used for the heat transfer measurement [1–3]. Some methods, while they have advanced the field, remain limited due to the complexity of implementing them in real applications. For example, optical methods such as interferometry [4,5] offer excellent spatial and temporal resolution, making it well-suited to observe rapid, localized thermal events. However, it requires complex setups and optical access, which can be difficult to achieve. The other methods, such as calorimetric [6–8] and acoustic methods [9,10], are generally suited to time-averaged measurements due to their slower transient response (generally a few milliseconds to seconds). Nevertheless, recent works show that advanced acoustic sensors combined with machine learning can extend the capabilities of such techniques to a higher frequency range [11,12]. The inverse methods [13,14] offer moderate transient performance but are prone to instability and reduced accuracy due to noise and rapid transients.

*Corresponding Author: himanshi.kharkwal@univ-fcomte.fr

To study boiling, four main types of methods are used. Among optical methods, one of the most used is infrared (IR) thermography, which combines input power measurements with spatial temperature mapping to estimate local heat flux. This technique offers non-intrusive, full-field temperature measurements and is widely used to visualize microlayer evaporation, hot spots and bubble–surface interactions. However, it depends heavily on accurate emissivity data, viewing angle and surface roughness. For instance, using IR Thermography, Može *et al.* [15] investigated the impact of surface morphology on boiling heat transfer and Xia *et al.* [16] used it in studying biphilic surfaces effect on pool boiling. The uncertainty on finding average surface temperature in these studies was 1°C and 1.5°C, respectively.

Other commonly used heat-transfer measurement techniques rely on thermal sensors. For instance, temperature sensors are used to infer heat flux through Fourier's law. Studies using this technique include the work from El-Genk and Pourghasemi [17], who used K-type thermocouples located 0.8 mm below the boiling surface. Similarly, Chuang *et al.*[18] and Jaswal *et al.* [19] used thermocouple arrays in copper blocks; Yuan *et al.* [20] used nine thermocouples in three planes for porous surfaces; and Li *et al.* [21] used platinum resistance thermometers (first thermometer being 3.81 mm below boiling surface) for cryogenic fluids. This method is appropriate for direct measurement and requires simple instrumentation. However, it provides only point measurements and therefore lacks the ability to detect the detailed lateral distribution of heat flux across the surface. Another technique of heat flux measurement uses integrated microheaters and sensors to infer heat flux using input voltage and current values. For instance, Hutter *et al.* [22] used nickel resistance temperature detectors and one microheater. Similarly, Liu *et al.* [23] used platinum RTDs and a Pt thin-film heater in pool boiling experiments. Although this technique enables direct measurement at a precise location, it requires sophisticated microfabrication. The characteristics obtained by the sensor depend on its overall geometry. Furthermore, some works involve using thermopile-based heat flux sensors. For instance, Barthes *et al.* [24] and Tadrist *et al.* [25] used commercial CAPTEC [26] sensors to study heat transfer during a single vapor bubble growth. This technique enables direct measurement of heat flux via a thermopile. However, the spatial resolution and temporal response are inherently limited by the geometry of the sensor and how it is integrated into the test surface. The direct measurement methods, particularly thin-film-based microfabricated sensors [27,28], strike a balance between spatial resolution and temporal response, as they can be optimized in terms of design, material and size to deliver robust and accurate measurements. In the direct method, differential gradient-based sensors are of interest due to their fast response time (typically on the order of micro-seconds to seconds) [29,30]. These sensors provide high-precision, real-time measurements, but are complex to fabricate and implement.

Overall, significant progress has been made in pool-boiling diagnostics; however, existing techniques are still limited in their ability to detect the rapid, localized thermal events associated with single-bubble dynamics. These limitations highlight the need for a miniaturized, thermally optimized sensor capable of controlled bubble nucleation and high-resolution measurement at the bubble footprint. The objective of this work is to investigate single-bubble heat transfer by developing sensing and metrological tools capable of capturing fast thermal dynamics with high temporal resolution. In the first part of the study, boiling meters based on commercial heat flux sensors were reviewed and tested, revealing their performance limitations in resolving rapid, localized boiling events. To overcome these limitations, a micro-boiling meter (μ-BM) was designed, fabricated, and evaluated. Although the full sensor package is millimeter-sized, the device is referred to as "micro" because its functional sensing films and membrane structure are microfabricated with micron-scale thicknesses and dimensions. Preliminary testing of the μ-BM was conducted in both radiative and pool boiling environments, enabling a comprehensive evaluation of the sensor's temporal performance.

## 2. MATERIALS AND METHODS

This work aims to develop a measurement device that can capture local wall temperature and heat flux signals during pool boiling experiments, particularly in single-bubble regimes and with varying surface inclinations. The device must be able to operate under both atmospheric and sub-atmospheric saturation conditions.

**2.1 Experimental setup.** The pool boiling experimental setup (see Fig. 1) consists of a cubic Teflon–Plexiglas cell that is filled to 75% of its volume with the test liquid. Two test fluids were used depending on the experiment. Either 3M™ Fluorinert™ FC-72 or Flutec PP1 was used. Flutec is an equivalent of FC-72 (that 3M has stopped producing and marketing). These two fluids have the same chemical formula and similar thermophysical properties. The test cell is equipped with two K-type thermocouples (liquid and vapor), a pressure transducer and a cryostat-controlled heat exchanger for thermal control. For heat flux and wall temperature measurement underneath bubble, boiling meters are used. Two of them integrate CAPTEC thermopile heat-flux sensors with a sensitivity of 2.82 ± 0.2 μV/W/m$^2$ and are detailed later on and on Fig.2. A single artificial nucleation site was fabricated directly on the top surface of the heat-flux sensor using an indentation method, with a diameter of 200 ± 20 μm and a depth of 150 ± 20 μm. Wall temperature is measured locally using a T-type thermocouple positioned 400 μm below the nucleation site. The boiling meter was mounted on a rotating shaft, enabling inclination studies to be conducted with ±0.5° accuracy.

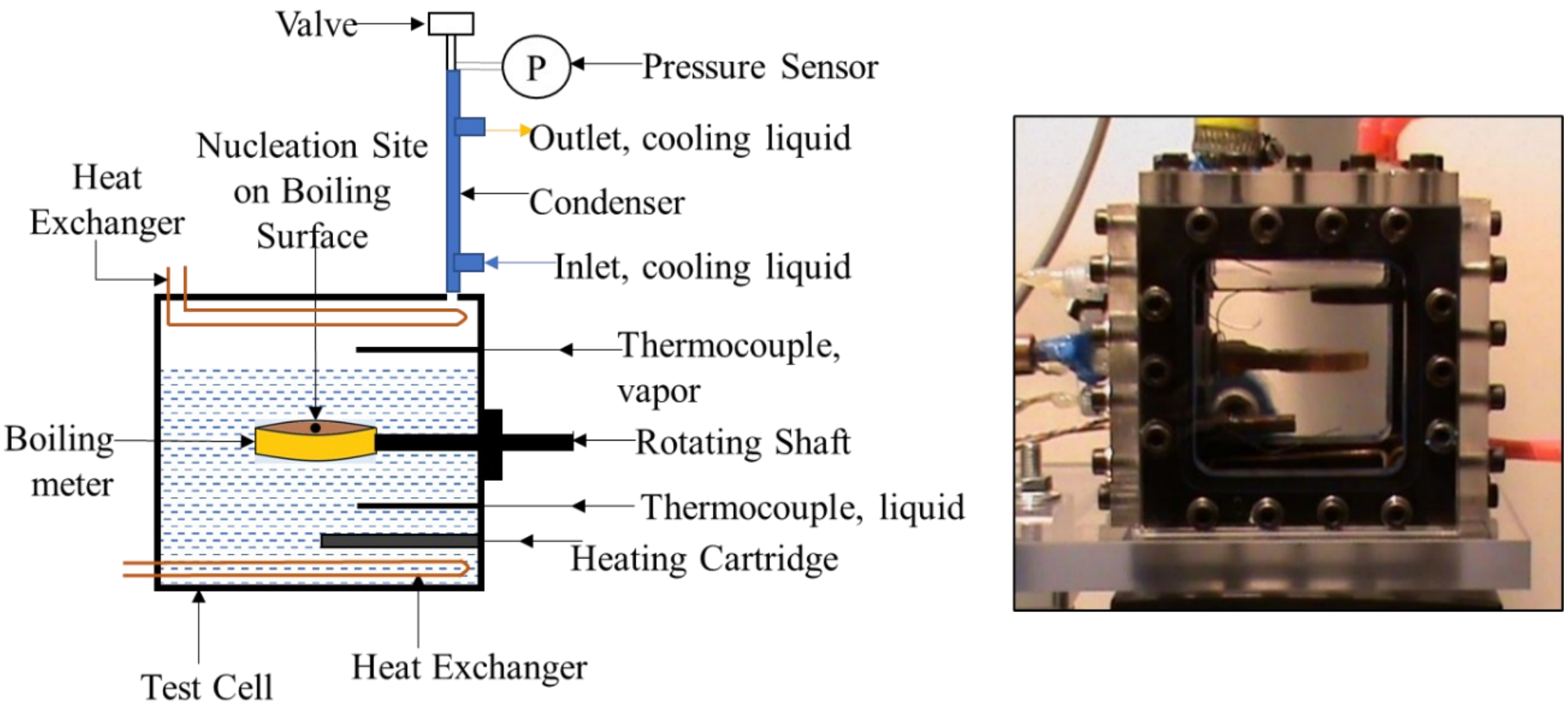

(a) (b)

**Fig. 1** (a) Schematic of the experimental setup showing the main components; (b) photograph of the experimental setup.

*Instrumentation.* The test cell (Fig. 1) is equipped with K-type thermocouples placed both in liquid and vapor, and a pressure sensor (SCX 15DN) that has ±1% pressure accuracy over wide temperature range (0°C to 85°C ). All the thermocouples were calibrated against a Pt-100 sensor, and have 0.2 °C type-B uncertainty in temperature measurements. Heating of the boiling meter is imposed with an uncertainty of ±0.005 W.

In the present work two boiling meters (BM-1 and BM-2) were used. Image and scheme of those two boiling meters are given on Fig.2. The BM-1, which was developed in previous work [31], consists of a central resistive heating film that is sandwiched between two 2 mm-thick copper pellets. Three CAPTEC [26] thermopile heat-flux sensors were integrated: one above the upper copper pellet, one below the lower pellet and one around the edge to quantify lateral heat transfer. A 5 mm resin layer was used to minimize radial heat losses and a 1 mm insulating layer was placed on the bottom. The device had dimensions of 30 mm in diameter and 6 mm in thickness. The heat-flux sensors comprise of distributed T-type thermocouples which provide total flux through the surface. A separate thermocouple, placed directly beneath the nucleation site, yields a local wall temperature. BM1 enabled detailed monitoring of the bubble cycle at low frequencies and large inclination angles; however, as it will be discussed in section 2.2, its response time was limited. To address this issue, BM-1 was redesigned and a second design (BM-2) was developed with reduced size. In BM-2, the copper pellets were reduced to 800 μm and relocated to the side that comes into contact with the liquid to facilitate surface preparation. The lateral heat-flux sensor was removed and the resin thickness was reduced to 1 mm. The overall size of BM-2 was reduced to 22 mm in diameter and 3.5 mm in thickness.

*Test Procedure.* The tests were conducted with the objective of ascertaining the maximum achievable frequency of the boiling meters. The working fluid was boiled for several hours to reach the target temperature and pressure. The boiling meter was then mounted upside down (180° orientation) and heated via the internal resistive heater, while nucleation was triggered at the designated site. Heat flux and wall temperature were recorded continuously using a data acquisition system. The surface inclination was subsequently adjusted to the next test angle, and measurements were repeated. Testing continued until the boiling meter reached its performance limit, defined as the point at which it could no longer track individual bubbles and only reported average heat-flux values.

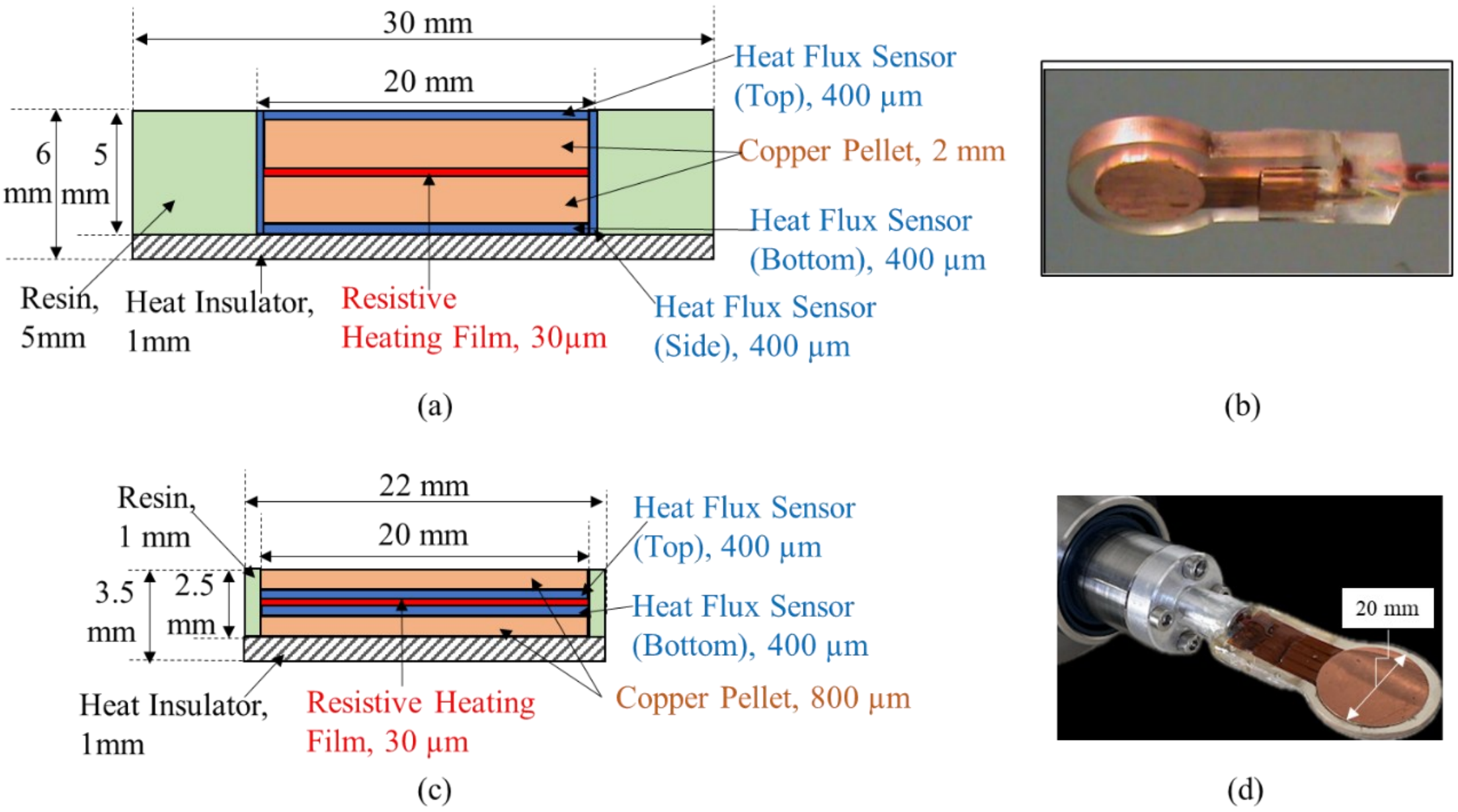

**Fig. 2** (a) Schematic and (b) photograph of Boiling Meter 1 (BM-1); (c) schematic and (d) photograph of Boiling Meter 2 (BM-2).

**2.2 Assessment of BM-1 and BM-2 Performance.** This section is limited to evaluating the dynamic measurement capability of the two boiling meters. The intention here is not to interpret boiling behavior (which can be found in Kharkwal *et al.* [32]), but rather to determine how each device responds to rapid thermal fluctuations. By comparing signals recorded at different bubble emission frequencies, we can identify the temporal limitations of BM-1 and BM-2, and establish the maximum frequencies at which individual bubble events can be resolved.

*BM-1 Performance.* Measurement results obtained with the BM1 for different angles of inclination of the nucleation surface are shown in Fig. 3. BM-1 successfully captured distinct heat flux and temperature fluctuations at low bubble emission frequencies (approximately 0.7 Hz), such as at 180° under sub-atmospheric pressure (see Fig. 3 a). At 180°, each period (oscillation) on the curve corresponds to a vapor bubble, from its nucleation to its detachment from the surface. However, slight changes in inclination (179°, Fig 3.b) increased the frequency of bubble departure, causing the sensor to lose the ability to track individual bubbles. The signal recorded corresponds to an average value for the phenomenon. The periodic peaks observed are related to the deactivation/activation of the nucleation site, and the findings related to BM-1 are discussed in detail in [32]. Similar averaging occurred at smaller angles (e.g., 30°, Fig 3.c), where the sizes and timescales of the bubbles were too small in relation to the BM-1's thermal inertia [32]. In order to improve the temporal response of the boiling meter, a second design of boiling meter BM-2 was created and tested.

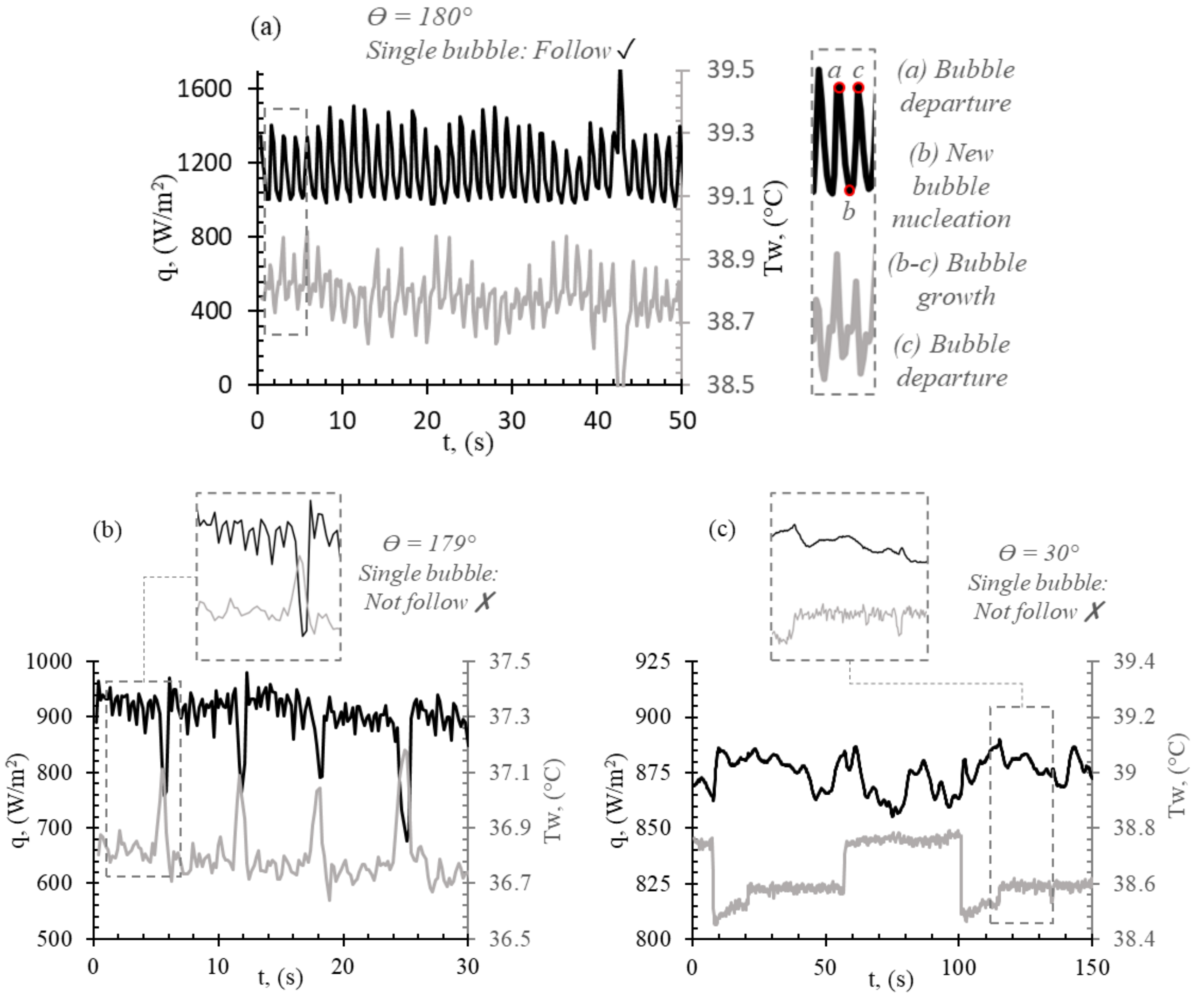

**Fig. 3** The heat flux (black) and wall temperature (grey) signals were recorded using Boiling Meter 1. (a) Case at 180°[32], where the sensor resolves each bubble. (b) Case at 179°, where the sensor no longer follows individual bubbles. (c) Case at 30°, where the sensor provides an averaged heat flux value, regardless of whether the nucleation site is active or inactive [31,32]. The working conditions were electrical heating power P = 0.46 W, fully degassed liquid at Tsat = 36.4 °C, and Psat = 0.47 bar.

<u>*BM-2 Performance.*</u> BM-2 was tested at surface inclinations of 175°, 170°, 150° and 130°. In these tests, the nucleation site emitted bubbles at two frequencies for each orientation: a slow frequency for large bubbles and a faster frequency for small bubbles. The figures for each case are shown on Fig. 4. Each oscillation corresponds here to one bubble, from its nucleation to its detachment.

- At 175°: BM-2 resolved both frequencies (0.8 Hz and 2.7 Hz).
- At 170°: frequencies of 0.7 Hz and 4.9 Hz were also fully captured.
- At 150°: the lower frequency (1 Hz) was clear, but the 5.5 Hz signal exhibited reduced amplitude.
- At 130°: lower frequency of 0.8 Hz was captured but the highest frequency (7 Hz) could not be tracked and the BM-2 output became averaged.

When compared with the results obtained with the BM-1, the BM-2 allows for more accurate and higher frequency bubble tracking. Nevertheless, although the BM-2 sensor can individually quantify the change in heat flux associated with the presence of each vapor bubble at relatively high angles/low emission frequencies, at lower angles/larger emission frequencies, the sensor sensitivity is no longer sufficient to detect individual bubbles. Thus, the BM-2 enable improving the temporal response, but the dynamics of higher-frequency bubbles (>5–6 Hz) remain poorly resolved. Therefore, a new boiling meter that uses microfabrication techniques has been developed to try to overcome these limitations. The microfabrication process is detailed in the following section.

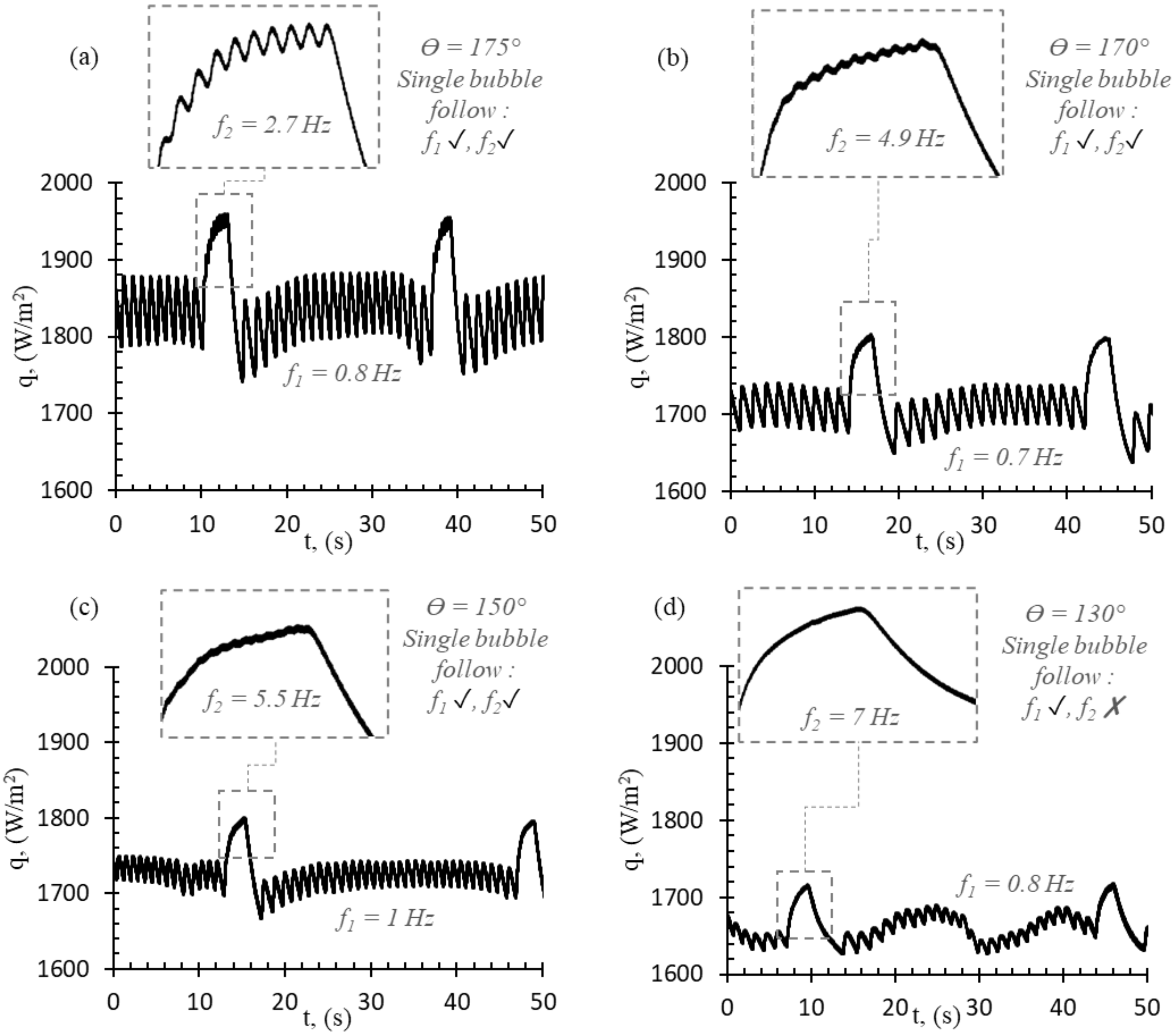

**Fig. 4** BM-2 heat-flux signals at various inclination angles at : (a) 175°: Both bubble frequencies are resolved. (b) 170°: Both frequencies are still captured. (c) 150°: The low frequency was resolved and the high frequency attenuated. (d) 130°: High-frequency bubbles are no longer tracked and the signal is averaged. The working conditions were electrical heating power P = 1W, partially degassed liquid at Tsat = 31.4 °C, Psat = 0.44 bar.

## 3. MICROFABRICATION OF MICRO-BOILING METER (μ-BM)

**3.1 Design Overview.** Given the limitations encountered with the commercial heat flux sensors used, we decided to design and micro-fabricate a heat flux sensor, with the aim of achieving a better compromise between response time and sensitivity. The literature shows that normal and tangential heat-flux sensors have similar temporal response ranges. Although tangential sensors, which are typically thermopile-based, can achieve higher sensitivity by increasing the number of thermocouple junctions, this also increases their size and slows their response [33,34]. By contrast, normal-gradient based heat-flux sensors (N-HFS) offer a more balanced performance, enabling sensitivity and response time to be optimized through appropriate material selection and geometric design [30].

*N-HFS based micro-boiling meter (μ-BM).* For the present study, the boiling meter must meet two essential requirements: (1) it must be capable of generating a single, controlled bubble and (2) it must provide sufficient sensitivity and temporal resolution to capture associated wall temperature and heat flux fluctuations. Accordingly, a detailed study combining a literature review, one-dimensional thermal calculations, an evaluation of thermal and mechanical parameters, and numerical simulations was conducted. Based on these analyses, a N-HFS based micro-boiling meter was designed (see Fig 5) which integrates four functional components: a normal gradient heat flux sensor composed of platinum resistance temperature detectors (Pt-RTDs) for high-resolution temperature measurement, a thin-film platinum heater for controlled heat input, and a nucleation site. Unlike conventional sensors, the RTD is directly exposed to the liquid without passivation layers, ensuring high temporal resolution and precise surface measurements.

Furthermore, heat flux is measured through the temperature gradient created across a micro-engineered substrate comprising three bonded layers: a Borofloat 33 (BF33) membrane; a silicon (Si) wafer containing a central air cavity with a micropillar; and a second thermally oxidized silicon wafer bonded using PermiNex® [35] resin. The air cavity and micropillar serve two purposes: they provide strong thermal insulation beneath the sensing region and they form the structural base for fabricating a cylindrical nucleation site aligned exactly at the sensor center.

Temperature is monitored on both sides of the substrate using two spiral platinum RTDs: RTD-top, which is positioned on the BF33 membrane to ensure thermal isolation and direct interaction with the bubble; and RTD-bot, which is patterned on the lower silicon wafer. To enable controlled bubble nucleation, a spiral platinum heater concentric with RTD-bot is integrated on the underside of the device. The dimensions and thicknesses of the thin-film elements were selected to ensure adequate temperature sensitivity, minimal self-heating, feasibility of fabrication, and sufficient power density for bubble generation. The final parameters are:

- RTD-top: 190 nm thick Ti/Pt, 10 µm track width, 1 mm spiral diameter
- RTD-bot + heater: 260 nm thick Ti/Pt, 20 µm track width and 1 mm spiral diameter.

Furthermore, a cylindrical shape was selected for the nucleation site based on literature and micro-fabrication feasibility. The diameter of the nucleation site (100 µm) was determined based on Hsu's criteria [36] and preliminary tests (with nucleation site fabricated using deep reactive-ion etching on a silicon wafer, and diameter ranging from 5 to 250 µm) conducted prior to the microfabrication of the N-HFS.

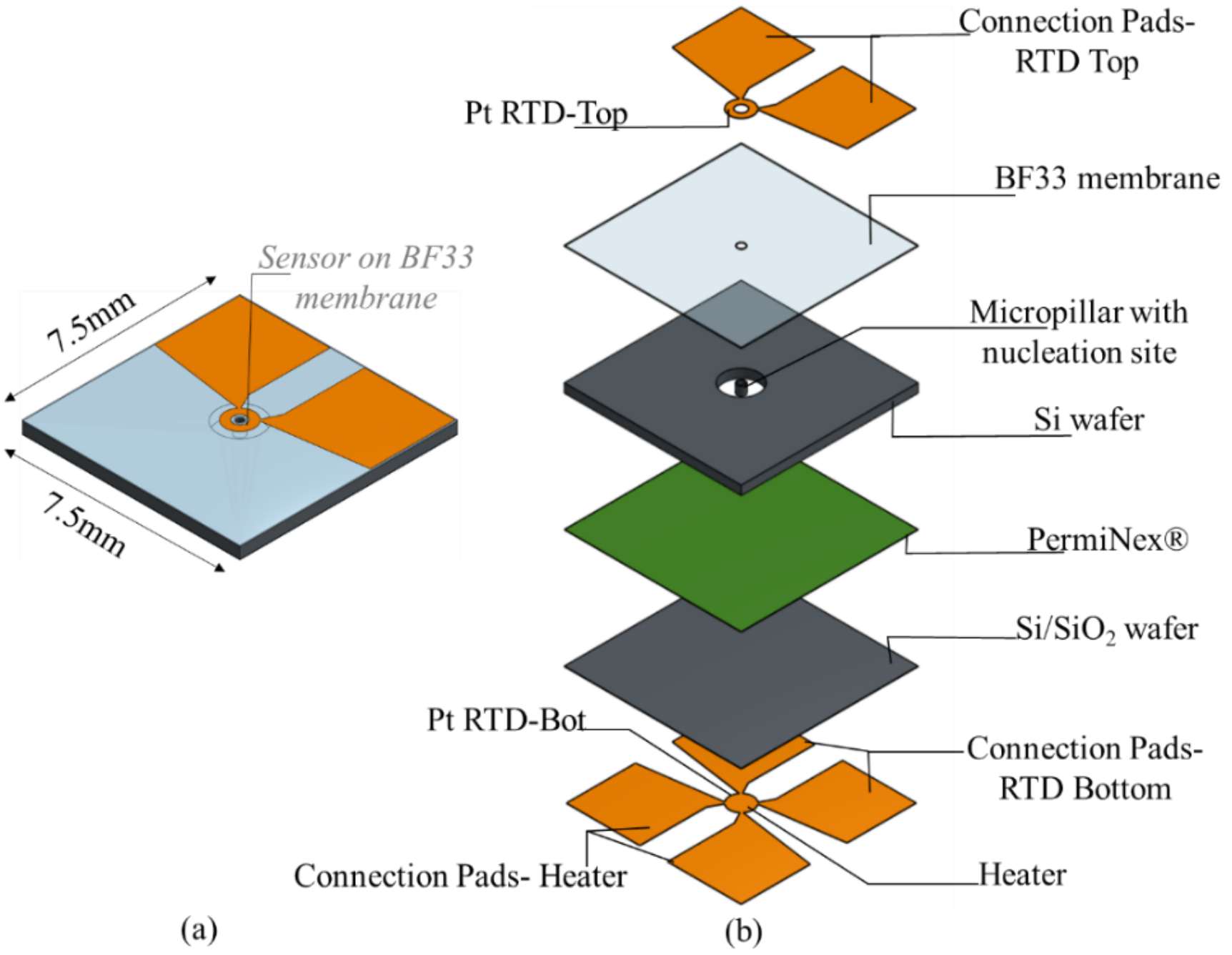

**Fig. 5** Schematic showing (a) the N-HFS with the top sensor (RTD-top) on a BF33 membrane, and (b) an exploded view of all components.

**3.2 Microfabrication of N-HFS**. The microfabrication was carried out in two parts. The top part involved the RTD-top on the BF33 membrane and the air-void silicon wafer. To do so, first a BF33 wafer was bonded to a silicon wafer using anodic bonding, then the BF33 was lapped and polished to a thickness of 40 µm. Photolithography was performed, followed by physical vapor deposition of Pt on a few-nanometer Ti adhesion layer to form the RTD-top. The air void (0.66 mm$^3$) was created on the silicon side using photolithography and deep reactive-ion etching (DRIE). Furthermore, the bottom part included RTD-bot fabrication on the thermally oxidized silicon wafer through photolithography and Ti/Pt deposition. After completing both parts, the top and bottom parts were bonded using PermiNex® adhesive bonding. Finally, the bonded assembly was diced, and a nucleation site was created at the central micropillar using laser ablation. Fig. 6.a shows the schematic of the microfabrication process of the N-HFS, and Fig. 6.b shows the fully fabricated micro-boiling meter highlighting the essential components: Pt RTDs (c and e), a heater (e) and a nucleation site in the center of the top platinum RTD (c and d).

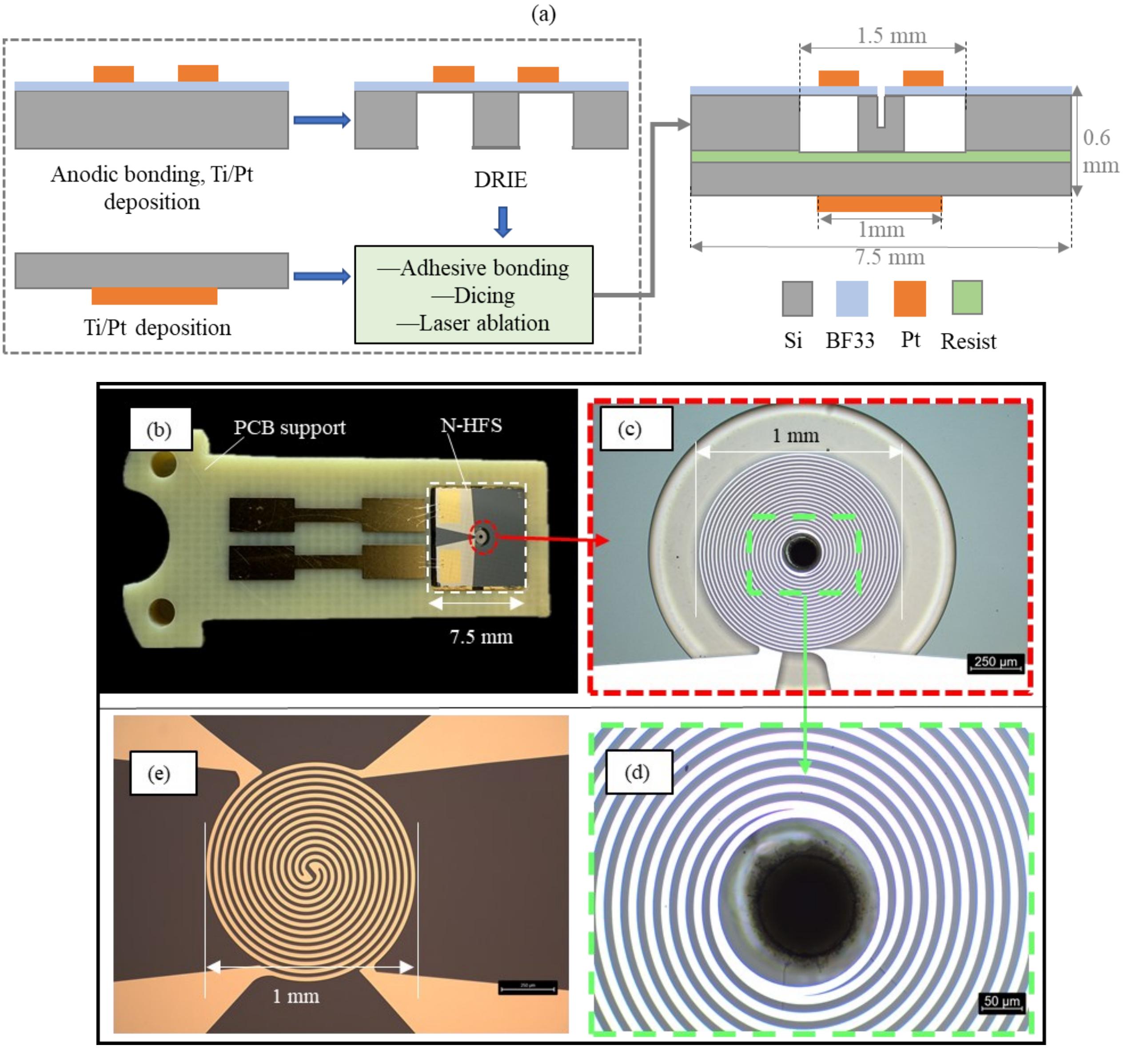

**Fig. 6** (a) Schematic showing microfabrication process of N-HFS, (b) Photograph of the complete device (µ-BM), with the 1 mm sensing area highlighted in red. (c) Top view of the sensing region showing the Pt-RTD patterned on the suspended glass membrane, encircling nucleation site. (d) Magnified view of the nucleation site located at the center. (e) Bottom view of the device illustrating the two Pt spirals corresponding to the Pt-RTD and Pt-heater.

## 4. RESULTS AND DISCUSSION

**4.1 Temporal Characterization of BM-2 and µ-BM in Radiative Environments.** The radiative response of both BM-2 and µ-BM was evaluated using a laser. A schematic of the experimental setup is shown in Fig. 7.a. From the obtained signals, to quantify the temporal response (not linear), the characteristic times were evaluated using the two parameters: $t_{rise}$ and $t_{fall}$. $t_{rise}$ is the time for the signal to increase from 10% to 90% of its total change and is related to the heating due to the laser (laser on). $t_{fall}$ is the time for the signal to decrease from 90% to 10% and corresponds to the natural cooling of the surface (laser off). The corresponding time constant was deduced for the worst-case scenario: maximum from ($t_{rise}$, $t_{fall}$) divided by ln(0.9/0.1).

BM-2 was tested using a 200 mW laser, with the beam expanded to a calculated diameter of 7 mm, determined from the divergence angle and the beam path length. The measured voltage was recorded using a Wheatstone bridge and an oscilloscope. The response of BM-2 is shown in Fig 7.e for the heating phase and Fig 7.f for the cooling phase. For the μ-BM, the RTD top was connected to a voltage divider circuit and the midpoint voltage was recorded using an NI acquisition system. This enabled real-time monitoring of resistance variations induced by laser heating. Multiple test conditions were applied by varying the laser power and beam diameter. The corresponding resistance–time traces are presented in Fig. 7(b–d). Specifically, the μ-BM was tested at a laser power of 100 mW with a beam diameter of 2 mm (Fig. 7.b) and 7mm (Fig 7. c). Then, the μ-BM was tested with a laser power of 200 mW and with beam diameters of 7 mm (Fig 7.d). In each case, the RTD-top exhibited a rapid increase in resistance during laser exposure, followed by a smooth decay once the pulse ended.

The characteristic rise and fall times for all test conditions are summarised in Table 1. For the μ-BM, the results (cases 1, 2 and 3 in Table 1) clearly show that the temporal response depends heavily on the power density of the incident beam. A higher power or a smaller beam diameter produces faster thermal transients, whereas a lower energy flux results in slower rise and fall times. This shows that the measured response time is affected not only by the sensor's intrinsic properties, but also by the thermal loading applied. Consequently, when reporting sensor time constant or response times, the excitation conditions must be explicitly stated, since temporal characteristics scale with the imposed heat flux.

For similar operating conditions (same laser power and spot size), the comparison between times obtained with BM-2 and μ-BM shows a real difference (cases 3 and 4 in Table 1): the times obtained with the μ-BM are more than 34 times shorter than those obtained with the BM-2. However, it is not appropriate to make a direct quantitative comparison because of differences in sensor size, material properties and optical absorption; but the trends suggest that the μ-BM exhibits a comparable faster response under similar irradiation levels. This enhanced performance can be attributed to its lower thermal mass, to localisation of the sensing element (Pt-RTD deposited on top of the surface) and the reduced heat-spreading volume provided by the membrane-based design.

**Table 1** Summary of radiative test conditions for the μ-BM and BM-2, including laser parameters and the corresponding characteristic rise and fall times.

| Case | Boiling meter | Laser power (mW) | Diameter of laser beam (mm) | $t_{rise}$ (ms) | $t_{fall}$ (ms) | time constant (ms) |
|---|---|---|---|---|---|---|
| 1 | μ-BM | 100 | 2 | 197 | 205 | 93 |
| 2 | μ-BM | 100 | 7 | 405 | 310 | 184 |
| 3 | μ-BM | 200 | 7 | 250 | 280 | 127 |
| 4 | BM-2 | 200 | 7 | 9325 | 9720 | 4423 |

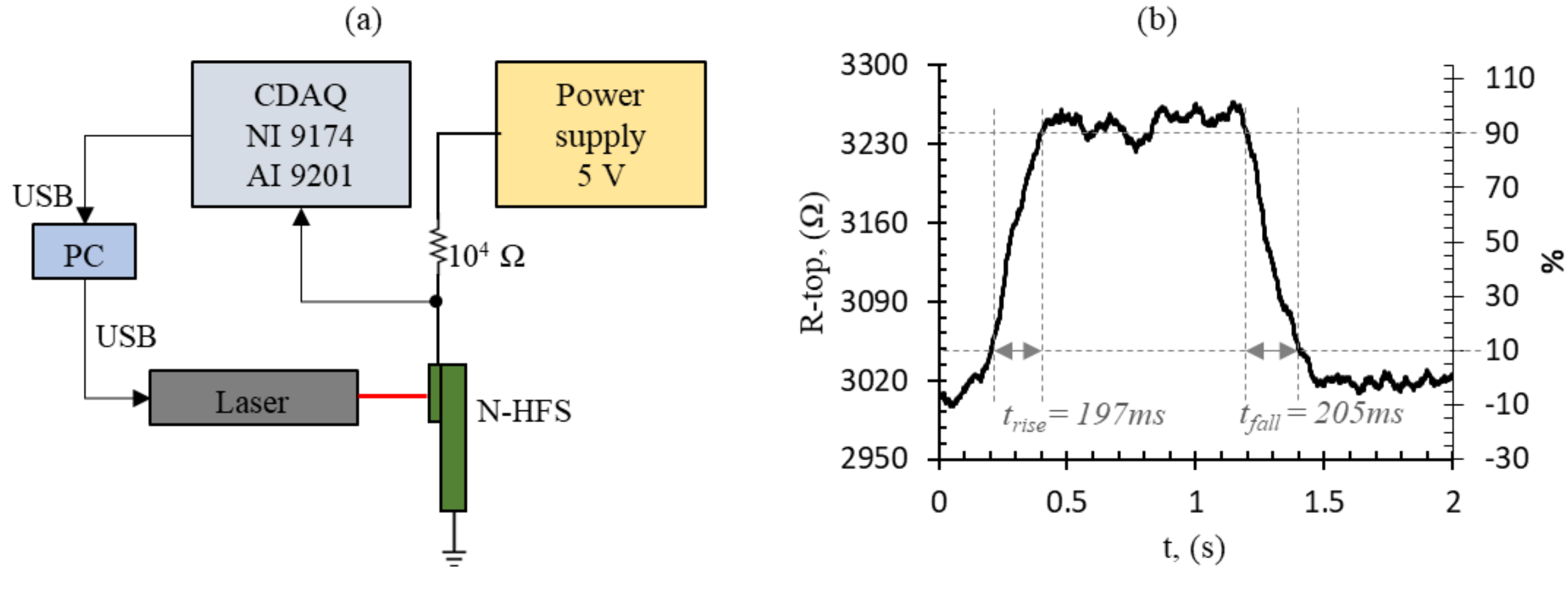

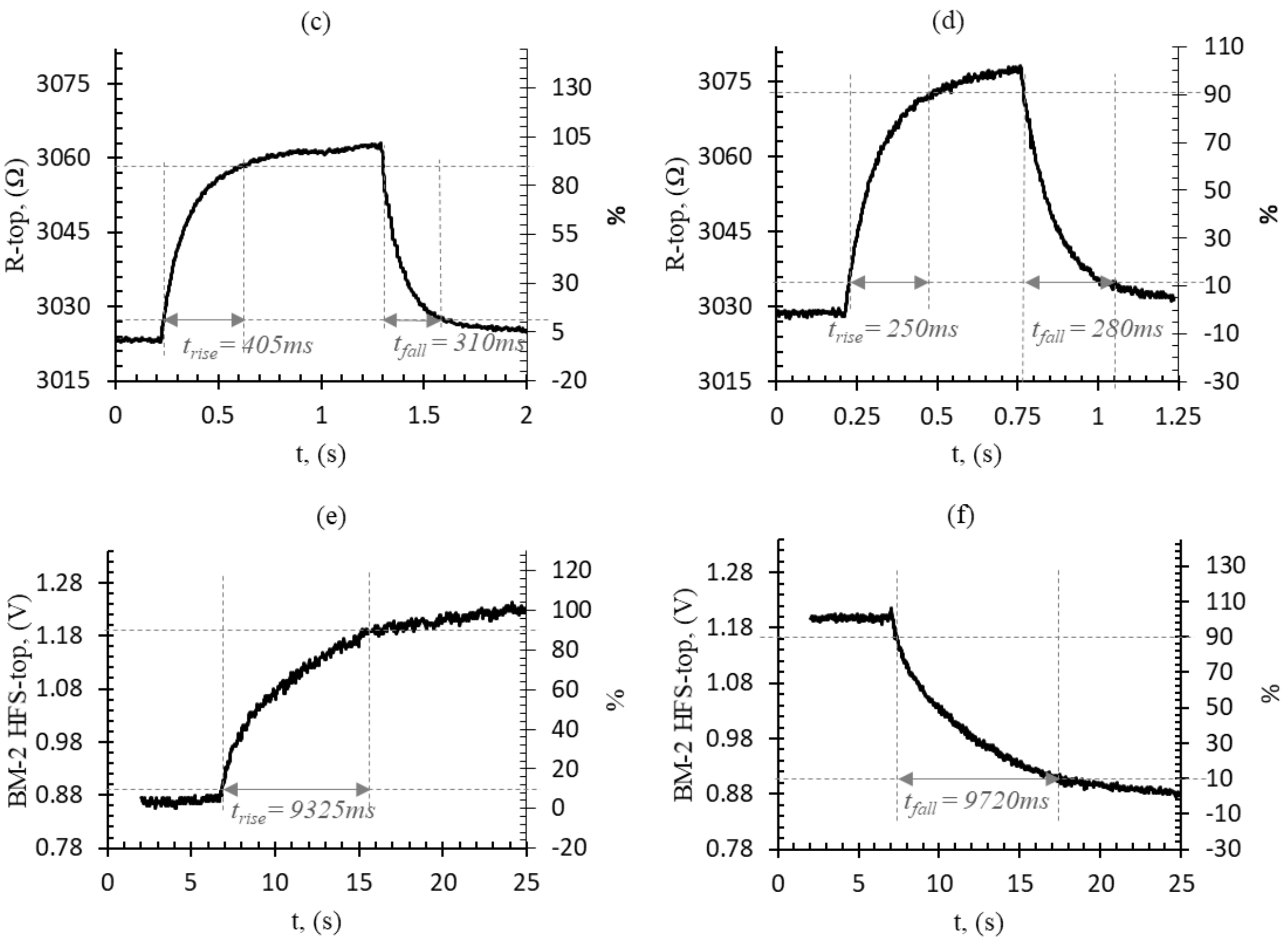

**Fig. 7** (a) Schematic of the experimental setup used for radiative testing. (b–d) Temporal response of the µ-BM under different radiative loading conditions: (b) 100 mW laser with 2 mm beam diameter, (c) 100 mW laser with 7 mm beam diameter, and (d) 200 mW laser with 7 mm beam diameter. (e–f) Temporal response of BM-2 under a 200 mW laser with 7 mm beam diameter, showing (e) the heating phase and (f) the cooling phase.

**4.2 Temporal Characterization of µ-BM in Pool Boiling.** To evaluate the temporal response and the sensitivity of the µ-BM, it was tested in pool-boiling conditions with Flutec PP1. Prior to testing, the fluid was boiled at atmospheric pressure (boiling temperature of about 57 °C) for two to three hours to remove dissolved air, resulting in a partially degassed liquid. The platinum heater on the µ-BM was then activated to initiate nucleation via Joule heating (heating power from 0.5 W to 0.6 W). To ease the onset of the nucleate boiling on the artificial nucleation site, initially, the nucleation surface was oriented at 180°. Temperatures from thermocouple in the test cell, values of Pt-RTD from top and bottom and video (fast camera) were recorded. As the Pt-RTD exhibits a linear relationship between resistance and temperature (resistances and thermocouples were calibrated on an AOIP bench in the laboratory), the resistance signal reflects the temperature fluctuation signature. For the present analysis, the focus is on the temporal response; therefore, only the resistance versus time data is examined.

The results are shown in Fig. 8, which presents the variation in resistance of the two RTDs during bubble emission at different frequencies. The black curve corresponds to the top RTD (which is in direct contact with the bubble, since it is created on the artificial site, located at the centre of the spiral-shaped RTD, cf. Fig.6-c and d) while the grey curve represents the bottom RTD (which is intertwined with the resistance serving as heating).

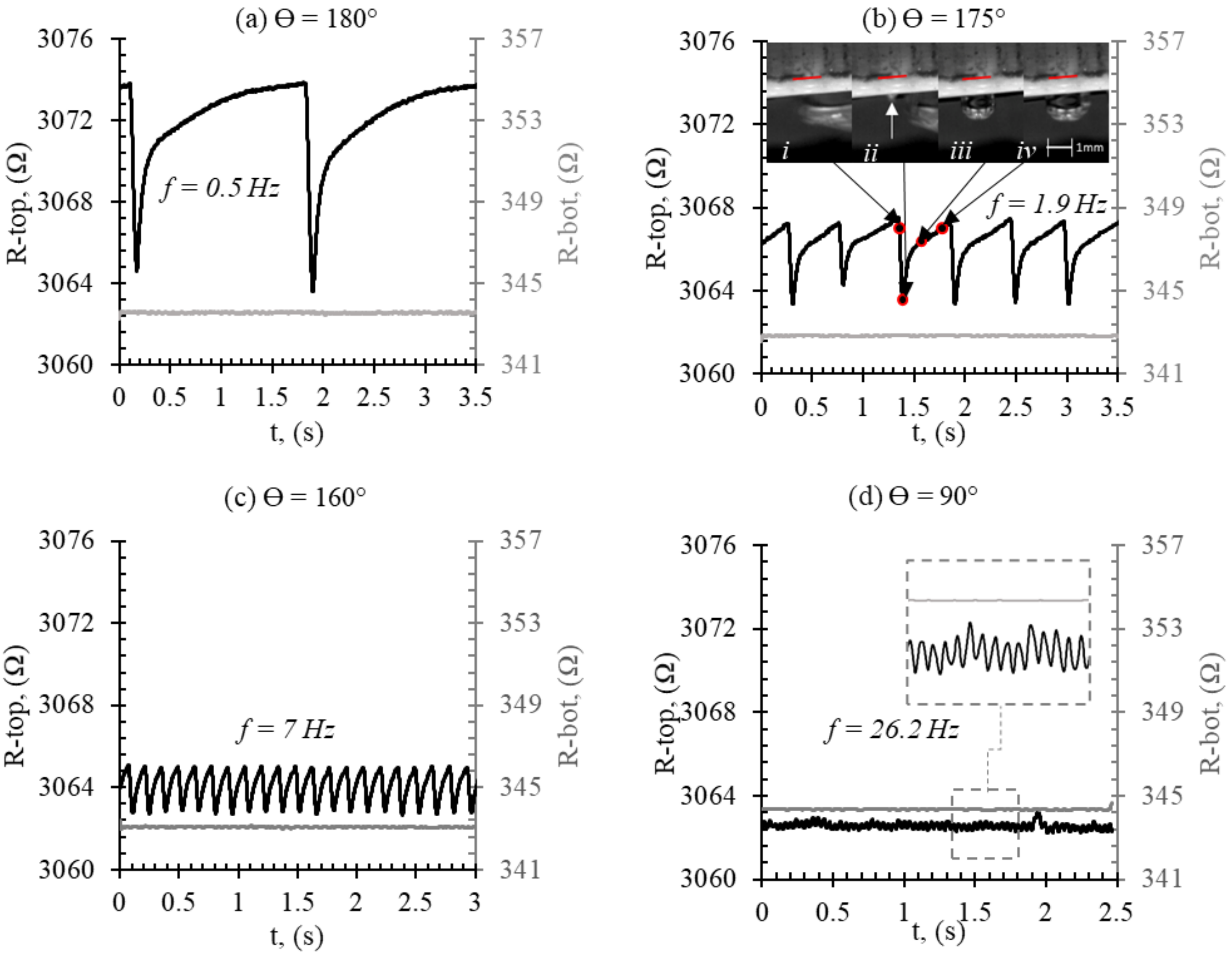

**Fig. 8** Resistance signals from RTD-top and RTD-bot during bubble emission. RTD-top captures distinct bubble frequencies at surface inclinations of (a) 180°, (b) 175°, (c) 160°, and (d) 90°. In (b), corresponding images of a single bubble cycle at 175° are shown, depicting: (i) bubble departure marked by a sharp drop in RTD-top resistance, (ii) nucleation of a new bubble, (iii) bubble growth, and (iv) subsequent departure. The red line indicates the sensor localization, which is 1 mm in diameter.

Four surface inclination angles are here presented: 180°, 175°, 160° and 90°. All those angles are with an uncertainty of ±1°. For all the tested angles, while the grey curve remains constant, the black curve exhibits oscillations: each of them corresponds to a vapor bubble, from its nucleation to its departure from the nucleation site. Details regarding the bubble cycle are given on Fig. 8.b on which the different parts of a period are related to the corresponding image of the bubble obtained from the fast camera. On the resistance signal (black curve), the departure of the bubble corresponds to a sharp drop in resistance (i), then the nucleation of a new bubble corresponds to an increase in the resistance value (ii). This value increases during the growth of the bubble (iii) until a new drop in the signal (iv) when the bubble detaches and leaves the nucleation site.

At an inclination of 180°, the bubble emission frequency is approximately 0.5 Hz (Fig. 8.a); at 175° (Fig. 8.b) and 160° (Fig. 8.c), this increases to around 1.9 Hz and 7 Hz respectively. All these frequencies are clearly resolved by the sensor, whereas at equivalent frequencies, the two previous boiling meters (BM-1 and BM-2) were unable to follow the bubble cycle. When the surface inclination is changed to 90° (Fig. 8.d), the bubble frequency increases significantly, reaching 26.2 Hz. The decrease in the amplitude of the signal is related to the size of the bubble, which decreases when the angle decreases too. Nevertheless, despite the smaller size of the bubbles and the reduced magnitude of the associated resistance changes, the signal still exhibits a detectable and periodic response at these higher frequencies: the sensor is thus fast and sensitive enough to detect the bubble influence.

The previous experiments were repeated with other angles varying between 180° (bubble created below the surface) and 0° (bubble created on the surface). In order to visualize the influence of the angle on the signal, Fig.

9(a) shows the RTD-top's resistance response (i.e., signal amplitude) over a single bubble event, from nucleation to growth and departure, using the resistance values. For clarity, time (the period) is adimensioned. Furthermore, Fig. 9(b) presents the same cycle on a logarithmic scale to highlight the smaller fluctuations observed at higher frequencies. The parameter ΔR-top represents the resistance change associated with one complete bubble cycle. At 180°, where the bubbles grow larger and detach more slowly, the resistance change is highest at around 10 Ω. At 175°, the bubble frequency increases and the corresponding ΔR-top decreases to around 4 Ω. At 160° and 150°, it reaches around 2 to 2.3 Ω. For 90° and 0°, where bubble generation is fastest (about 20Hz with our operating conditions) and bubble size is smallest, ΔR-top is reduced to 0.3–0.4 Ω. Even with these smaller amplitudes, the μ-BM clearly captures the bubble-induced fluctuations, demonstrating its capability to resolve bubble dynamics over a wide range of frequencies.

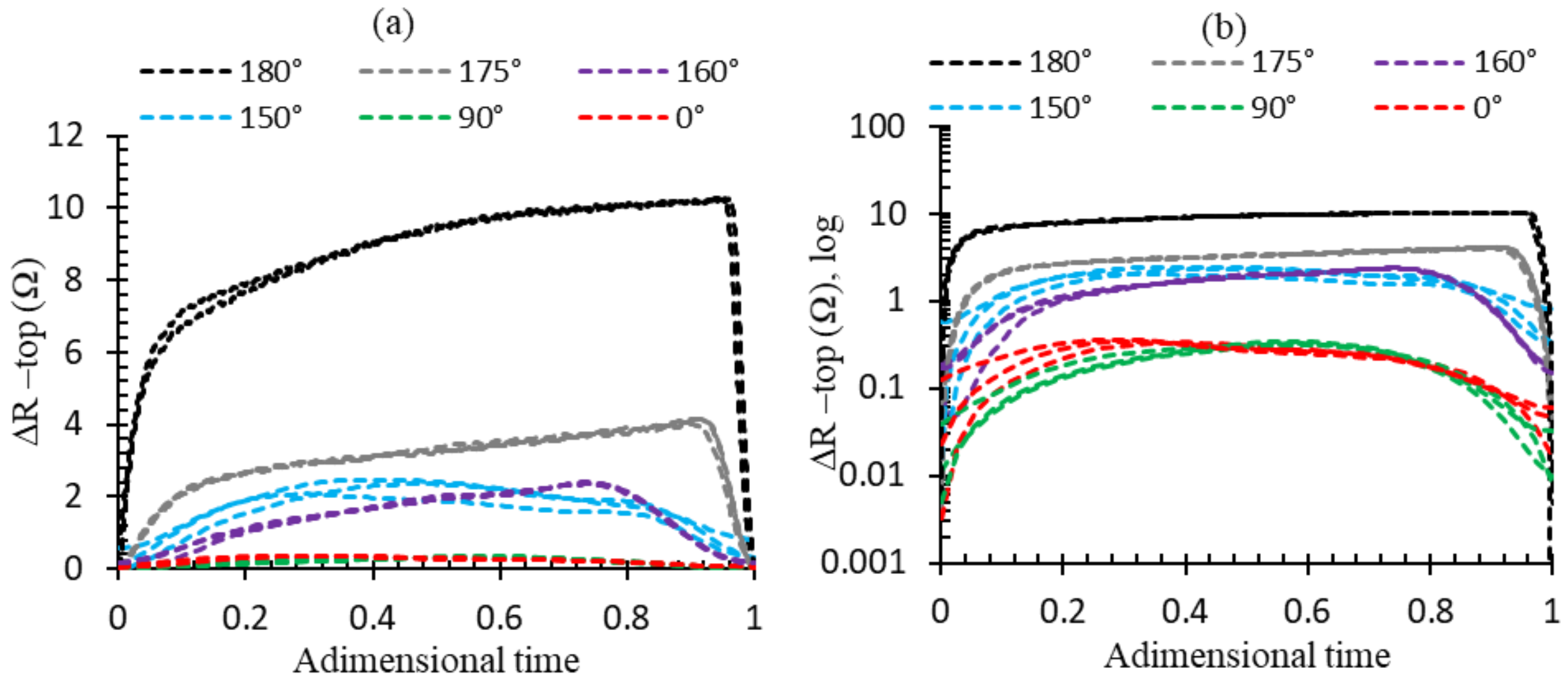

**Fig. 9** Change in RTD-top resistance over single bubble cycle—from nucleation to departure—shown in (a) linear scale and (b) logarithmic scale.

## 5. CONCLUSION

This paper presents the development of a microfabricated boiling meter (μ-BM), which is designed to measure wall temperature and heat flux fluctuations associated with single-bubble dynamics. The limitations of earlier devices (BM-1 and BM-2) were first identified. Although these devices were effective at low bubble frequencies, they were unable to resolve higher-frequency events, which motivated the development of a new sensor architecture. The resulting design methodology combined thermal analysis, mechanical constraints and microfabrication feasibility to create a normal heat flux sensor that integrates a BF33 membrane and an air cavity substrate with a central micropillar to define a nucleation site. The sensor also incorporates two thin-film platinum resistance temperature detectors (Pt RTDs) with an integrated heater. Radiative excitation tests have shown that, on the one hand, the microfabricated μ-BM has shorter characteristic times than the BM-2, which incorporates commercial sensors. Furthermore, the μ-BM has faster rise and fall times under higher incident power densities, confirming that its temporal response is strongly influenced by the applied thermal load and not solely by the sensor design. Pool boiling tests have demonstrated that the μ-BM can detect bubble cycles from ~0.5 Hz to ~26 Hz, while previously tested sensors (BM-1 and BM-2) did not have sufficient sensitivity and/or response time to track the bubble cycle at smaller angles and/or frequencies above 7 Hz. The μ-BM provides localized, high-resolution detection at the bubble footprint and is a promising tool for advancing the understanding of the physics of individual bubbles. Further work will focus on calibration of the μ-BM (in order to accurately determine the heat flux), followed by a systematic study of the influence of various parameters, including inclination, on heat exchange and growth dynamics.

# ACKNOWLEDGMENTS

The authors thank the French National Research Agency (ANR) for its financial support of the TraThI project (ANR-21-CE50-0009-01 grant). This work was carried out within the framework of the EIPHI graduate school (ANR-17-EURE-0002 grant). This work was supported by the French RENATECH network and its FEMTO-ST technology center.

# NOMENCLATURE

| | | | | | |
|---|---|---|---|---|---|
| $f$ | frequency | (Hz) | $t_{rise}$ | characteristic rise time | (ms) |
| $q$ | heat flux | (W/m$^2$) | $t_{fall}$ | characteristic fall time | (ms) |
| $R$ | resistance | ($\Omega$) | | | |